\documentclass[%
reprint, 
superscriptaddress,
nofootinbib,
amsmath,amssymb,
aps,
prd,
showkeys]{revtex4-2}

\usepackage{graphicx}% Include figure files
\usepackage{dcolumn}% Align table columns on decimal point
\usepackage{bm}% bold math
\usepackage{hyperref}% add hypertext capabilities
\usepackage{cleveref}

\usepackage{comment}

\hypersetup{
citecolor=blue,
 colorlinks=true,
 linkcolor=blue,
 filecolor=magenta,
 urlcolor=cyan,
}

\usepackage[dvipsnames]{xcolor}
\usepackage{soul}
\usepackage[normalem]{ulem}
\usepackage{xspace}
\usepackage{booktabs}
\usepackage{multirow}
\usepackage[normalem]{ulem}
\usepackage{fancyhdr}
\usepackage{graphicx,amsfonts,amssymb,amsbsy}
\usepackage{amsmath,amsthm,latexsym}
\usepackage[utf8]{inputenc}  
\usepackage{aas_macros}
\usepackage{soul}
\usepackage{tabularx}
\usepackage{enumitem}

\setlist[itemize]{leftmargin=20pt, itemsep=1pt}%Default: leftmargin=25pt, itemsep=4pt

\begin{document}

\preprint{APS/123-QED}

\title{Towards a unified hadron-quark equation of state for neutron stars within the relativistic mean-field model}

\author{Marcos O. Celi}
\email[]{mceli@fcaglp.unlp.edu.ar}
\affiliation{Grupo de Astrofísica de Remanentes Compactos, Facultad de Ciencias Astronómicas y Geofísicas, Universidad Nacional de La Plata, Paseo del Bosque S/N, La Plata (1900), Argentina}
\affiliation{CONICET, Godoy Cruz 2290, Buenos Aires (1425), Argentina}

\author{Mauro Mariani}
\email[]{mmariani@fcaglp.unlp.edu.ar}
\affiliation{Grupo de Astrofísica de Remanentes Compactos, Facultad de Ciencias Astronómicas y Geofísicas, Universidad Nacional de La Plata, Paseo del Bosque S/N, La Plata (1900), Argentina}
\affiliation{CONICET, Godoy Cruz 2290, Buenos Aires (1425), Argentina}

\author{Milva G. Orsaria}
\affiliation{Grupo de Astrofísica de Remanentes Compactos, Facultad de Ciencias Astronómicas y Geofísicas, Universidad Nacional de La Plata, Paseo del Bosque S/N, La Plata (1900), Argentina}
\affiliation{CONICET, Godoy Cruz 2290, Buenos Aires (1425), Argentina}

\author{Ignacio F. Ranea-Sandoval}
\affiliation{Grupo de Astrofísica de Remanentes Compactos, Facultad de Ciencias Astronómicas y Geofísicas, Universidad Nacional de La Plata, Paseo del Bosque S/N, La Plata (1900), Argentina}
\affiliation{CONICET, Godoy Cruz 2290, Buenos Aires (1425), Argentina}

\author{Germán Lugones}
\email[]{german.lugones@ufabc.edu.br}
\affiliation{Universidade Federal do ABC, Centro de Cîencias Naturais e Humanas,
Avenida dos Estados 5001- Bangú, CEP 09210-580, Santo André, SP, Brazil.}

\date{\today}% It is always \today, today,

\begin{abstract}
The equation of state of dense matter remains a central challenge in astrophysics and high-energy physics, particularly at supra-nuclear densities where exotic degrees of freedom like hyperons or deconfined quarks are expected to appear. Neutron stars provide a unique natural laboratory to probe this regime. In this work, we present EVA--01, a novel equation of state that provides a unified description of dense matter by incorporating both hadron and quark degrees of freedom within a single relativistic mean-field Lagrangian, from which the equation of state is derived at finite temperature. The model extends the density-dependent formalism by introducing a Polyakov-loop-inspired scalar field to dynamically govern the hadron-quark phase transition, following the approach of chiral mean-field models.
The resulting model is consistent with a wide range of theoretical and observational constraints, including those from chiral effective field theory, massive pulsars, gravitational-wave events, and NICER data. We analyze its thermodynamic properties by constructing the QCD phase diagram, identifying the deconfinement, chiral, and nuclear liquid-gas transitions. As a first application, we model the evolution of proto-neutron stars  using isentropic snapshots and explore the implications of the slow stable hybrid star hypothesis. Our findings establish EVA--01 as a robust and versatile framework for exploring dense matter, bridging the gap between microphysical models and multimessenger astrophysical observations.
\end{abstract}

%Use showkeys class option if keyword
\maketitle

\section{Introduction}\label{sec:intro}

The composition of matter at the extreme densities found within the cores of neutron stars (NSs) remains a key unsolved problem in modern physics and astrophysics. 
Below nuclear saturation density, the equation of state (EoS) is well constrained by terrestrial experiments and chiral Effective Field Theory (cEFT) calculations~\cite{Hebeler:2013eos, Drischler:2020hwd}. Moreover, at asymptotically high densities, restrictions from perturbative QCD (pQCD) are available~\cite{Annala:2020efq}. Unfortunately, in the intermediate regime relevant for studying NS interiors, no first-principles description (or experimental facilities) is available. For this reason, effective models that capture (some of) the essential features of dense matter must be used. This regime is precisely where a transition from hadronic matter to deconfined quark matter is expected to occur.

A common strategy for constructing a hybrid EoS is to ``paste'' independent models for the hadronic and quark phases. While sophisticated descriptions exist for each sector individually, this two-phase construction introduces an artificial separation and requires additional assumptions to connect the two regimes. Most notably, the hadron–quark transition is typically imposed through a Maxwell or Gibbs construction, rather than arising directly from the underlying theory. An alternative and more physically robust strategy is to develop a unified framework in which both hadrons and quarks are described within a single effective Lagrangian. Recently, unified descriptions based on quarkyonic matter and parity-doublet dynamics have been proposed, in which quark degrees of freedom coexist with partial confined matter at intermediate densities \citep{Bikai:2025qmw}. In other models, like the chiral mean-field (CMF) approach~\cite{Dexheimer:2010ana, Kumar:2024mna}, inspired by the Polyakov-loop extended Nambu–Jona-Lasinio (PNJL) model~\cite{Fukushima:2003fw}, a scalar field $\Phi$ is introduced to mediate deconfinement, dynamically suppressing quark degrees of freedom at low densities and temperatures. 

Motivated by this last approach, we introduce a similar $\Phi$-dependent mechanism into the density-dependent relativistic mean-field (DDRMF)  parametrization SW4L~\cite{Spinella:2019hns, Malfatti:2020dba, Celi:2024doh}, leading to a unified hadron–quark model named EVA--01, the first Extended VAriable-coupling relativistic mean-field model that unifies hadronic and quark phases within a single Lagrangian. This construction allows for a consistent description of both sectors, preserving the nuclear phenomenology of the hadronic interaction while extending it to include quark degrees of freedom within the same field-theoretical structure.

To be physically meaningful, the EVA--01 EoS is parameterized to be consistent with a wide range of constraints. From a microphysical standpoint, these include chiral Effective Field Theory (cEFT) calculations applicable in the range of $n_0 \lesssim n \lesssim 2n_0$~\cite{Hebeler:2013eos, Drischler:2020hwd} and perturbative QCD (pQCD) at asymptotically high densities~\cite{Annala:2020efq}.  Furthermore, the EoS must comply with modern astrophysical observations, including gravitational waves from binary mergers like GW170817~\cite{Abbott:2017goo, Abbott:2017gwa, Abbott:2017moo}, and mass-radius measurements from high-mass pulsars~\cite{Demorest:2010ats, Antoniadis:2013amp, Cromartie:2020rsd} and NICER X-ray timing~\cite{Riley:2019anv, Miller:2019pjm, Riley:2021anv, Miller:2021tro}.

The very nature of the hadron-quark phase transition, which EVA--01 is designed to model, remains a subject of intense theoretical debate and investigation. In the high-density, low-temperature regime relevant for NSs, the transition has long been modeled as first-order. Within this framework, its characteristics are strongly influenced by the hadron-quark surface tension: a high value leads to a sharp, discontinuous transition via the Maxwell construction~\cite{Endo:2011roh,Wu:2018eoq}, while a low value results in a Gibbs construction with an extended mixed-phase region, potentially featuring complex geometrical structures known as hadron-quark ``pasta''~\cite{ju:2021hqp,mariani:2024qhp}. However, some recent studies have challenged this first-order picture, arguing that a smooth crossover behavior might occurs also in the cold, dense regime~\cite{Schafer:1999coq, Masuda:2013hqc, Hirono:2019qhc, Fujimoto:2025shq, Fukushima2025qpd}.

Beyond this first-order versus crossover dichotomy, more intricate scenarios have been proposed. These include the emergence of intermediate phases, such as quarkyonic matter or the \textit{spaghetti of quarks with glueballs} phase, where quarks are partially deconfined while gluons remain in a confined state~\cite{Fujimoto:2025nso}. Other possibilities involve the formation of inhomogeneous phases, for instance, through interweaving chiral spirals~\cite{Fukushima:2024tpd}.

These diverse theoretical perspectives underscore that the nature of the quark-hadron transition at high density is far from settled. Nevertheless, the occurrence of such a transition, regardless of its specific form, is expected to have profound and observable consequences for the structure of compact stars. The most direct implication is the potential existence of hybrid stars, compact objects harboring a deconfined quark matter core.

The astrophysical manifestation of a first-order phase transition within a hybrid star depends critically on the conversion dynamics at the hadron-quark interface. The outcome is determined by the interplay between the hadron-quark conversion timescale and the characteristic timescale of the star's fundamental radial oscillation mode. Two limiting regimes are commonly considered: the {\it rapid} conversion scenario, in which phase conversion is much faster than the perturbation timescale, and the {\it slow} scenario, in which the conversion proceeds on a timescale much longer than the star's oscillation period. In the rapid conversion limit, it has been shown that the loss of dynamical stability occurs at the maximum-mass configuration, as indicated by the standard condition $\partial M/\partial \varepsilon_c = 0$ ~\cite{Pereira:2018pte}. Conversely, slow conversions can lead to the formation of an extended branch of stable hybrid configurations~\cite{Pereira:2018pte}, known as slow stable hybrid stars (SSHSs) \cite{lugones:2023ama}, even in the region where stars would typically be considered unstable ($\partial M/\partial \varepsilon_c < 0$)~\cite{mariani:2019mhs, Tonetto:2020dgm, rodriguez:2021hsw, Goncalves:2022ios,Mariani:2022omh,Ranea:2022bou,Ranea:2023auq,Ranea:2023cmr,Rau:2023tfo,Rau:2023neo,Rather:2024roo,Mariani:2024cas,laskos:2025xja}.

In this work, we apply the EVA--01 model in two primary investigations. We begin by constructing the phase diagram predicted by EVA--01, analyzing the hadron-quark deconfinement and chiral symmetry restoration transitions across a wide temperature-density range. This provides a baseline for evaluating the behavior of the model at both zero and finite temperature. We then turn to a specific astrophysical application, studying the thermal evolution of hybrid proto-neutron stars (PNSs), focusing on the slow conversion scenario. This scenario is chosen as it represents the more general case, since its set of stable solutions encompasses all configurations that are stable under the rapid conversion limit. To capture the essential thermodynamic features of this process while avoiding the complexities of a full simulation with neutrino transport~\cite{pons:1999eop}, we adopt the simplified scheme of isentropic snapshots with fixed lepton fractions, an approach widely used to address PNS evolution~\cite{Prakash:2001eoa, lattimer:2004tpo, Shao:2011eop, Mariani:2017ceh, Ghosh:2024etm}.

This paper is organized as follows. In Section~\ref{sec:model}, we describe the EVA–01 model in detail. Section~\ref{sec:results} presents the main results of our work, including the phase diagram of the model and the astrophysical analysis of PNSs. In Section~\ref{sec:conclus}, we summarize our key findings and discuss their relevance in the modern context of NSs astrophysics.

\section{Unified Hadron-Quark EVA-01}\label{sec:model}

We base our unified hadron-quark model on the SW4L parametrization, a nonlinear DDRMF model widely used to describe dense hadron matter in astrophysical contexts \citep{Malfatti:2020dba, Mariani:2022omh, Celi:2024doh}. To extend the DDRMF model to include quark degrees of freedom, we follow the strategy proposed by \citet{Dexheimer:2010ana}, originally developed for the SU(3) nonlinear $\sigma$ model (see also Refs. \citep{Kumar:2024mna, Celi:2025etr} for modern implementations). This construction is conceptually similar to that of the Polyakov-loop-extended Nambu-Jona-Lasinio (PNJL) model (see, for example, Refs.~\citep{Fukushima:2008pdi,Malfatti:2019hqm} and references therein), where a scalar field $\Phi \in [0, 1]$ is introduced as an order parameter that drives the transition from the confined phase to the deconfined one.

Building on these ideas, we develop a novel unified model for dense matter: the Extended VAriable-coupling relativistic mean field (EVA-01) model. Its Lagrangian density contains contributions from baryons (nuclear matter, $n,p$), hyperons ($\Lambda^0, \Sigma^+, \Sigma^0, \Sigma^-, \Xi^0,$ $\Xi^-$), the four states of the $\Delta$ particle ($\Delta^-,\Delta^0,\Delta^+,\Delta^{++}$), quarks ($u, d, s$), leptons ($e^-, \mu^-, \nu_e$), scalar ($\sigma, ~\sigma^*$), vector ($\omega,~ \phi$), and isovector ($\rho$) meson fields, and the scalar field, $\Phi$, associated with deconfinement.

The complete EVA-01 Lagrangian density reads
\begin{equation}\label{eq:lagrangian}
\mathcal{L}= \mathcal{L}_b+ \mathcal{L}_q+\mathcal{L}_{\sigma\omega\rho}+\mathcal{L}_{NL\sigma}+\mathcal{L}_{\phi\sigma^*}+ \mathcal{L}_l - U_\Phi \, ,
\end{equation}
where the baryon and quark terms are given, respectively, by~\cite{Malfatti:2020dba,Celi:2024doh, Kumar:2024mna},
\begin{eqnarray}
\label{eq:intb}
  \mathcal{L}_b &=& \sum\limits_b \overline\psi_b\bigl[\gamma_{\mu}
    (i\partial^{\mu}-g_{\omega b}\omega^{\mu}-g_{\phi b}\phi^{\mu}-\tfrac{1}{2}g_{\rho b}\boldsymbol{\tau}\cdot
    \boldsymbol{\rho}^{\mu})\nonumber\\ &&-(m_b-g_{\sigma
      b}\sigma-g_{\sigma^*
    b}\sigma^*+g_{\Phi b}\Phi^2)\bigr]\psi_b \, ,
\end{eqnarray}
and
\begin{eqnarray}
\label{eq:intq}
  \mathcal{L}_q &=& \sum\limits_q \overline\psi_q\bigl[\gamma_{\mu}
    (i\partial^{\mu}-g_{\omega q}\omega^{\mu}-g_{\phi q}\phi^{\mu}-\tfrac{1}{2}g_{\rho q}\boldsymbol{\tau}\cdot
    \boldsymbol{\rho}^{\mu})\nonumber\\ &&-(m_q-g_{\sigma
      q}\sigma-g_{\sigma^*
      q}\sigma^*+g_{\Phi q}(1-\Phi)\bigr]\psi_q \, .
\end{eqnarray}
The sum over $b$ runs over all baryons, including $\Delta$ resonances. The coefficients $g_{ib}$ (see Table~\ref{table:couplings_and_masses}) denote the meson-baryon coupling constants, inherited from the DDRMF-SW4L parametrization \cite{Spinella:2017asi, Spinella:2019hns, Malfatti:2020dba, Celi:2024doh}. The sum over $q$ includes the three quark flavors considered in the model: $u$, $d$, and $s$. Analogously, $g_{iq}$ refers to the meson-quark coupling constants, which have been tuned to reproduce their vacuum masses while ensuring that the resulting EoS remains compatible with astrophysical NS constraints. These values are also listed in Table~\ref{table:couplings_and_masses}.

The mesonic Lagrangian terms read \cite{Malfatti:2020dba,Celi:2024doh},
\begin{eqnarray}
\mathcal{L}_{\sigma\omega\rho}&=&\tfrac{1}{2}\left(\partial_{\mu}\sigma\partial^{\mu}\sigma-m^2_{\sigma}\sigma^2\right)-\tfrac{1}{4}\omega_{\mu\nu}\omega^{\mu\nu}+\tfrac{1}{2}m^2_{\omega}\omega_{\mu}\omega^{\mu}\nonumber\\&&
  -\tfrac{1}{4}\boldsymbol{\rho}_{\mu\nu}\cdot\boldsymbol{\rho}^{\mu\nu}
  +\tfrac{1}{2}m^2_{\rho}\boldsymbol{\rho}_{\mu}\cdot\boldsymbol{\rho}^{\mu} \, , %\nonumber,
\end{eqnarray}
\begin{eqnarray}
\mathcal{L}_{NL\sigma}&=&-\tfrac{1}{3}\tilde{b}_{\sigma}m_n\left(g_{\sigma
    N}\sigma\right)^3-\tfrac{1}{4}\tilde{c}_{\sigma}\left(g_{\sigma
    N}\sigma\right)^4 \, , 
    \label{eq:NLsigma} %\nonumber,
\end{eqnarray}
\begin{eqnarray}
  \mathcal{L}_{\phi\sigma^*} &=& -\tfrac{1}{4}\phi^{\mu\nu}\phi_{\mu\nu}+\tfrac{1}{2}m^2_{\phi}\phi_{\mu}
  \phi^{\mu}\nonumber\\&&+\tfrac{1}{2}\left(\partial_{\mu}
  \sigma^*\partial^{\mu}\sigma^* -m^2_{\sigma^*}\sigma^{*2}\right) \, . %\nonumber .
\end{eqnarray}
In particular, Eq.~\eqref{eq:NLsigma} introduces nonlinearity through higher-order self-interaction terms for the $\sigma$ meson. As in the DDRMF-SW4L model, the isovector meson $\rho$ has a density-dependent coupling constant to account for medium effects. Using the results from Ref.~\cite{Spinella:2017asi}, these effects are parametrized as a function of the baryonic number density as follows :
\begin{equation}
g_{\rho i}(n_B)=g_{\rho i}\left(n_0\right) \exp \left[-a_\rho\left(\frac{n_B}{n_0}-1\right)\right] \, ,
\label{eq:rho_func}
\end{equation}
where $i$ refers to both baryons ($b$) and quarks ($q$). The values of the constants $\tilde{b}_\sigma$, $\tilde{c}_\sigma$, $a_\rho$, and the meson masses can be found in Refs.~\citep{Malfatti:2020dba, Celi:2024doh}. The functional form in Eq.~\ref{eq:rho_func} allows us to control the slope of the symmetry energy, an essential feature that helps satisfy the modern astrophysical constraints of NSs, without affecting other well-constrained properties of symmetric nuclear matter. For a more detailed discussion related to density dependent coupling constants in the context of RMF models and their impact on NS astrophysics, see, for example, Refs.~\cite{typel:1999rmf,Typel:2018rmf}. 

The hadron sector is calibrated to reproduce the nuclear matter properties at saturation density that are listed in Table~\ref{table:properties}.This is performed in order to agree with both experimental and first-principle calculations (see, for example, Refs. \cite{Burgio:2020teo, Choi:2021con, Oertel:2017eos} and references therein).

\begin{table*}[t!]
\setlength{\tabcolsep}{6pt}
\begin{tabular}{lcccccc}
\toprule
Particle      & $g_{\sigma j}$ & $g_{\omega j}$ & $g_{\rho j}$ & $g_{\sigma^* j}$ & $g_{\phi j}$ & Bare Mass [MeV] \\
\midrule
% Baryons
$n$           & $9.810$           & $10.391$          & $7.818$         & $0.000$             & -$3.547$         & $939.6$ \\
$p$           & $9.810$           & $10.391$          & $7.818$         & $0.000$             & -$3.547$         & $938.3$ \\
$\Lambda^0$   & $7.461$           & $8.253$           & $7.818$         & $1.924$             & -$6.333$         & $1115.7$ \\
$\Sigma^{+,0,-}$ & $5.441$        & $8.253$           & $7.818$         & $1.924$             & -$6.333$         & $1189.4$, $1192.6$, $1197.4$ \\
$\Xi^{0,-}$   & $5.900$           & $6.115$           & $7.818$         & $7.725$             & -$9.119$         & $1314.9$, $1321.3$ \\
$\Delta^{++,+,0,-}$      & $10.791$          & $11.430$          & $7.818$         & $0.000$             & -$3.547$         & $1232$ \\
\midrule
% Quarks
$u$           & $3.000$           & $1.500$           & $0.000$         & $0.000$             & $0.000$         & $2.16$ \\
$d$           & $3.000$           & $1.500$           & $0.000$         & $0.000$             & $0.000$         & $4.67$ \\
$s$           & $0.000$           & $1.500$           & $0.000$         & $3.000$             & $0.000$         & $93.4$ \\
\bottomrule
\end{tabular}
\caption{Dimensionless coupling constants of mesons with baryons ($g_{ib}$) and quarks ($g_{iq}$), and the corresponding bare masses used in the EVA-01 model.  For the $\Sigma$ and $\Xi$ isospin multiplets, the values in the last column are the bare masses corresponding to the charge states listed in the first column.}
\label{table:couplings_and_masses}
\end{table*}

\begin{table}[t]
\begin{center}
\begin{tabularx}{0.3\textwidth}{>{\centering\arraybackslash}X >{\centering\arraybackslash}X}
\toprule 
\multicolumn{2}{c}{Saturation Properties EVA-01} \\
\midrule
$n_0$~(fm$^{-3}$)    & $0.150$  \\
$E_0$~(MeV)          & $-16.00$ \\
$K_0$~(MeV)          & $250.0$ \\
${m_{N}}^*/m_N$      & $0.70$ \\
$S_0$~(MeV)          & $30.3$ \\
$L_0$~(MeV)          & $46.5$ \\ 
\bottomrule
\end{tabularx}
\caption{Nuclear matter properties at saturation density ($n_0$) for the hadronic sector of the EVA-01 model. The table lists the binding energy per nucleon ($E_0$), incompressibility ($K_0$), effective nucleon mass ($m_N^*/m_N$), symmetry energy ($S_0$), and its slope parameter ($L_0$).}
\label{table:properties}
\end{center}
\end{table}

The leptons are treated as a non-interacting Fermi gas, described by the standard Dirac Lagrangian summed over all relevant species ($l = e^-, \mu^-, \nu_{e^-}$):
\begin{eqnarray}
  \mathcal{L}_l =\sum_l\bar{\Psi}_l\left(i \gamma_\mu \partial^\mu-m_l\right) \Psi_l \, .
\end{eqnarray}

The Polyakov-like potential used in this work is inspired by the original Polyakov potential from the PNJL model~\citep{Fukushima:2008pdi}, including terms that not only depend on the temperature but also explicitly depend on the baryon chemical potential. As a result, the scalar field $\Phi$ affects thermodynamic quantities even in the low-temperature, high-density regime relevant for cold NSs. We adopt the functional form proposed by~\citet{Kumar:2024mna}:
\begin{align}
U_\Phi =& \left(a_0 T^4+a_1 \mu_B^4+ a_2 T^2 \mu_B^2\right) \Phi^2+\nonumber\\
&a_3 T_0^4 \ln \left(1-6 \Phi^2+8 \Phi^3-3 \Phi^4\right) \, .
\end{align}
The parameters $a_0$, $a_1$, $a_2$, $a_3$, and $T_0$ were adjusted to reproduce a physically reasonable phase diagram consistent with both QCD and NSs astrophysical constraints; their values are shown in Table~\ref{table:polyconstants}. As shown in Eqs.~(\ref{eq:intb}) and (\ref{eq:intq}), the scalar field $\Phi$ contributes to the effective masses, regulating the presence/absence of quarks and hadrons~\cite{Dexheimer:2010ana, Kumar:2024mna}:
\begin{align}
& m_b^*=m_{b}- g_{\sigma b} \sigma-g_{\sigma^* b} \sigma^*+g_{\Phi b} \Phi^2 \, , \label{eq:eff_mass_bar} \\
& m_q^*=m_{q}- g_{\sigma q} \sigma-g_{\sigma^* q} \sigma^*+g_{\Phi q}(1-\Phi) \, . \label{eq:eff_mass_qua}
\end{align}
The quantities $m_b$ and $m_q$ are the bare baryon and quark masses, respectively. Since leptons do not interact via the strong force, their masses remain unaffected by the field $\Phi$. The coupling constants $g_{\Phi i}$, with $i=b,q$, should be chosen to be large enough to ensure that when $\Phi \simeq 0$, baryon masses are low while quark masses are high, and conversely when $\Phi \simeq 1$, the opposite holds. Consequently, the presence of quarks will be favored for $\Phi \simeq 1$ and suppressed when $\Phi \simeq 0$, and vice versa for hadrons. 
Hence, we tuned the values of $g_{\Phi q}$ and $g_{\Phi b}$ to enforce a sharp separation between a pure hadron phase at low densities and a pure quark phase at high densities. This choice prevents any significant \textit{cross-contamination}\footnote{The term \textit{cross-contamination} denotes the coexistence of hadron and quark degrees of freedom. In this work, such an overlap is intentionally precluded by our parametrization, but in alternative applications of the model aimed at describing a mixed or crossover phase, this coexistence could be an intended physical feature.}. The selected values are listed in Table~\ref{table:polyconstants}.

After minimizing the mesonic fields, the full system of equations of motion reads:
\begin{eqnarray}
\label{eq:motion}
m_\sigma^2 \sigma & = &\sum_{i=b,q} g_{\sigma i} n_i^s -\tilde{b}_{\sigma}m_n g_{\sigma N}^3 \sigma^2-\tilde{c}_{\sigma}g_{\sigma N}^4\sigma^3  \nonumber \\
m_\omega^2 \omega & = &\sum_{i=b,q} g_{\omega i} n_i  \nonumber\\
m_\rho^2 \rho & = &\sum_{i=b,q} g_{\rho i} I_{3 i} n_i  \nonumber\\
m_{\sigma^*}^2 \sigma^* & = &\sum_{i=b,q} g_{\sigma^* i} n_i^s \\
m_\phi^2 \phi & =&\sum_{i=b,q} g_{\phi i} n_i  \nonumber \\
0 = &-&\sum_{b}2 g_{\Phi b} n_b^s\Phi+\sum_{q} g_{\Phi q} n_q^s \nonumber\\
&-&\left(a_0 T^4+a_1 \mu_B^4+a_2 T^2 \mu_B^2\right)\Phi \nonumber \\ 
&-&a_3 T_0^4 \frac{12 \Phi}{3 \Phi^2-2 \Phi-1} \nonumber,
\end{eqnarray}
where $I_{3i}$ is the third component of the isospin for each particle. Meson masses are listed in Ref.~\cite{Celi:2024doh}.

Each particle species (baryons, quarks, leptons) contributes to thermodynamic quantities through the following finite-temperature Fermi integrals. The scalar and vector densities, $n_i^s$ and $n_i$, are given by~\cite{Schmitt:2010dmi}:
\begin{equation}
\begin{aligned}
n_i^s & =\gamma_i \int \frac{d^3 p}{(2 \pi)^3}\left[f_{i-}(p)+f_{i+}(p)\right] \frac{m_i^*}{E_i^*}, \\
n_i & =\gamma_i \int \frac{d^3 p}{(2 \pi)^3}\left[f_{i-}(p)-f_{i+}(p)\right],
\end{aligned}
\label{Eq:ns_and_nb}
\end{equation}
with $i=b,q,l$. The expression for $n_i^s$ in Eq.~\eqref{Eq:ns_and_nb} corrects the sign error in Eq.~(14) of Ref.~\cite{Malfatti:2019hqm}, that becomes relevant only at temperatures $T \gtrsim 80$~MeV. The pressure of each species is given by~\cite{Schmitt:2010dmi}:
\begin{equation}
    P_i= \frac{\gamma_i}{3} \int \frac{d^3 p}{(2 \pi)^3} \frac{p^2}{E_i^*}\left[f_{i-}(p)+f_{i+}(p)\right] \, .
\end{equation}
To include thermal effects, we define the entropy density $\tilde{s}_i$~\footnote{We denote the entropy density by $\tilde{s}$ to distinguish it from the entropy per baryon, defined as $s \equiv \tilde{s}/n_B$.} as in Ref.~\cite{Malfatti:2020mdq}:
\begin{multline}
\tilde{s}_i = \frac{\gamma_i}{3} \int \frac{d^3 p}{(2 \pi)^3} \frac{p^2}{E_i^*} \Biggl\{ \left[ f_{i-}(p) - f_{i-}(p)^2 \right] \left( \frac{E_i^* - \mu_i^*}{T^2} \right) \\
+ \left[ f_{i+}(p) - f_{i+}(p)^2 \right] \left( \frac {E_i^* + \mu_i^*}{T^2} \right) \Biggr\} \, .
\end{multline}

In the preceding integrals, $\gamma_i$ is the degeneracy factor for each particle species, which accounts for spin and color degrees of freedom. Specifically, we use $\gamma_b = 2$ for spin-$1/2$ baryons, $\gamma_\Delta = 4$ for the $\Delta$ resonances, $\gamma_q = 6$ for quarks, and $\gamma_l = 2$ for leptons. The functions $f_{i\mp}$ are the Fermi-Dirac distributions for particles ($-$) and antiparticles ($+$), given by
\begin{equation}
f_{i \mp}(p)=\frac{1}{\exp \left[\frac{E_i^*(p) \mp \mu_i^*}{T}\right]+1} \, . 
\end{equation}

For baryons and quarks, $i=b,q$, $E^*_i$ stands for the effective energy given by
\begin{equation}
E_i^*(p)=\sqrt{p^2+m_i^{* 2}} \, ,
\end{equation}
where the effective masses $m_i^*$ are given by Eqs.~\eqref{eq:eff_mass_bar} and \eqref{eq:eff_mass_qua}; the dynamical chemical potential, $\mu_i^*$, is expressed as
\begin{equation}
    \mu_i^*= \mu_i-g_{\omega i} \omega-g_{\rho i}\rho I_{3i}-g_{\phi i}\phi-\widetilde{R} \, .
\end{equation}
The term
\begin{eqnarray}
\widetilde{R} =\sum_{i=b,q} \frac{\partial g_{\rho i}(n_B)}{\partial n_B}
I_{3i} n_i \rho  , %\nonumber
\label{eq:rear}
\end{eqnarray}
is the \textit{rearrangement} term, which arises from the density dependence of the coupling constants and ensures thermodynamic consistency~\cite{Hofmann:2001aot}. 

Given that leptons are non-interacting particles, their effective energy, mass, and chemical potential coincide with the corresponding free values: \mbox{$E_l^*=E_l$}, $m_l^*=m_l$ ($m_e=0.5$~MeV and $m_\mu=105.6$~MeV) and $\mu^*_l=\mu_l$.

To construct the total unified hybrid EoS, we calculate all the thermodynamic quantities by including contributions from all particle species and the Polyakov-like potential. Defining $b_i$ as the baryon charge, $b_i=1$ for baryons and $b_i=1/3$ for quarks, the total baryon number density, $n_B$, is given by
\begin{equation}
\label{eq:numdens}
n_B = \sum_{i=b,q} b_in_i  +n_{\Phi},
\end{equation}
where
\begin{equation}
    n_{\Phi} = -\frac{\partial U_\phi}{\partial \mu_B} = - 4 a_1 \mu_B^3 \Phi^2.
\end{equation}
Since $U_\Phi$ depends explicitly on $\mu_B$, this last term must be included to ensure thermodynamic consistency \cite{Peterson:2022cse}. Note that, as $\Phi\simeq0$ when quarks are absent, $n_{\Phi}$ becomes relevant only in the quark-dominated phase (i.e., $\Phi\simeq1$).

The total pressure, which also includes contributions from the nonlinear scalar term  $\mathcal{L}_{N L \sigma}$, the rearrangement term $\tilde{R}$ due to density-dependent couplings, and the potential $U_\Phi$, is given by
\begin{equation}
\begin{aligned}
P = &\sum_{i = b,q,l} P_i -\frac{1}{2} m_\sigma^2 \bar{\sigma}^2+\frac{1}{2} m_\omega^2 \bar{\omega}^2+\frac{1}{2} m_\rho^2 \bar{\rho}^2 \\
& -\frac{1}{3} \tilde{b}_\sigma m_N\left(g_{\sigma N}(n) \bar{\sigma}\right)^3-\frac{1}{4} \tilde{c}_\sigma\left(g_{\sigma N}(n) \bar{\sigma}\right)^4 \\
&+n_B \tilde{R} - U_\Phi. 
\end{aligned}
\end{equation}

The total entropy density reads
\begin{equation}
    \tilde{s} = \sum_{i=b,q,l} \tilde{s}_i + \tilde{s}_\Phi \, ,
\end{equation}
with the Polyakov-related contribution given by \cite{Peterson:2022cse}
\begin{equation}
        \tilde{s}_{\Phi} = -\frac{\partial U_\phi}{\partial T} = - \left(4a_0 T^3 + 2a_2 T \mu_B^2\right)\Phi .
\end{equation}
As before, this term becomes relevant only in thermodynamic regimes where quark matter is present, i.e., for $\Phi\simeq1$.

The total energy density is obtained through the thermodynamic Euler relation:
\begin{equation}
    \epsilon = -P+T\tilde{s}+\sum_{i=b,q,l}\mu_in_i+\mu_Bn_\Phi \, .
\end{equation}
Note that the term $\mu_Bn_\Phi$ must be included for thermodynamic consistency, due to the explicit dependence of $U_\Phi$ on $\mu_B$.

Additionally, the matter composition in NSs must satisfy the conditions of charge neutrality and \mbox{$\beta$-equilibrium}, with conservation of both electric charge and baryon number. The $\beta$-equilibrium imposes the following relation for chemical potentials of baryons and quarks
\cite{pons:1999eop, Prakash:2001eoa}:
\begin{equation}
\label{eq:chemeqb}
\mu_{i} = b_i\mu_B + q_i\,\mu_L \, ,
\end{equation}
where $q_i$ is the electric charge of $i$ species. Depending on the astrophysical scenario, we will either consider trapped electron neutrinos (with no muons), in which case  \mbox{$\mu_L= \mu_e - \mu_{\nu_e}$}, or the untrapped case with muons present, for which \mbox{$\mu_L= \mu_e = \mu_\mu$} (see next section for details). Charge neutrality is imposed by requiring:
\begin{equation}
\sum_{i=b,q,l} q_i\, n_i= 0 \, .
\label{eq:charge_0}
\end{equation}

The set of equations and conditions described above fully defines the EVA--01 model. We now proceed to apply this unified framework to construct the QCD phase diagram and to analyze relevant astrophysical scenarios. The results of these applications, along with further details on the parametrization choices, are presented in the following section.

\begin{table}[t!]
\begin{tabular}{c}
\toprule 
\quad $a_0 =-5.5$ , \quad $a_1 = -0.95 \times 10^{-3}$ \quad \\
\quad $a_2 = -88.69 \times 10^{-3}$ , \quad $a_3 = -0.396$ \quad \\
\quad $T_0 = 200$~MeV \quad \\
\quad $g_{\Phi q} = 2125$~MeV , \quad $g_{\Phi b} = 3 g_{\Phi q}$ \quad \\
\bottomrule
\end{tabular}
\caption{Parameters of the potential $U_\Phi$ and coupling constants $g_{\Phi q}$, and $g_{\Phi b}$ for the EVA-01 model.}
\label{table:polyconstants}
\end{table}

\section{Application of EVA-01}\label{sec:results}

%FIG. 1
\begin{figure}[t!]
    \centering
    \includegraphics[width=1\linewidth]{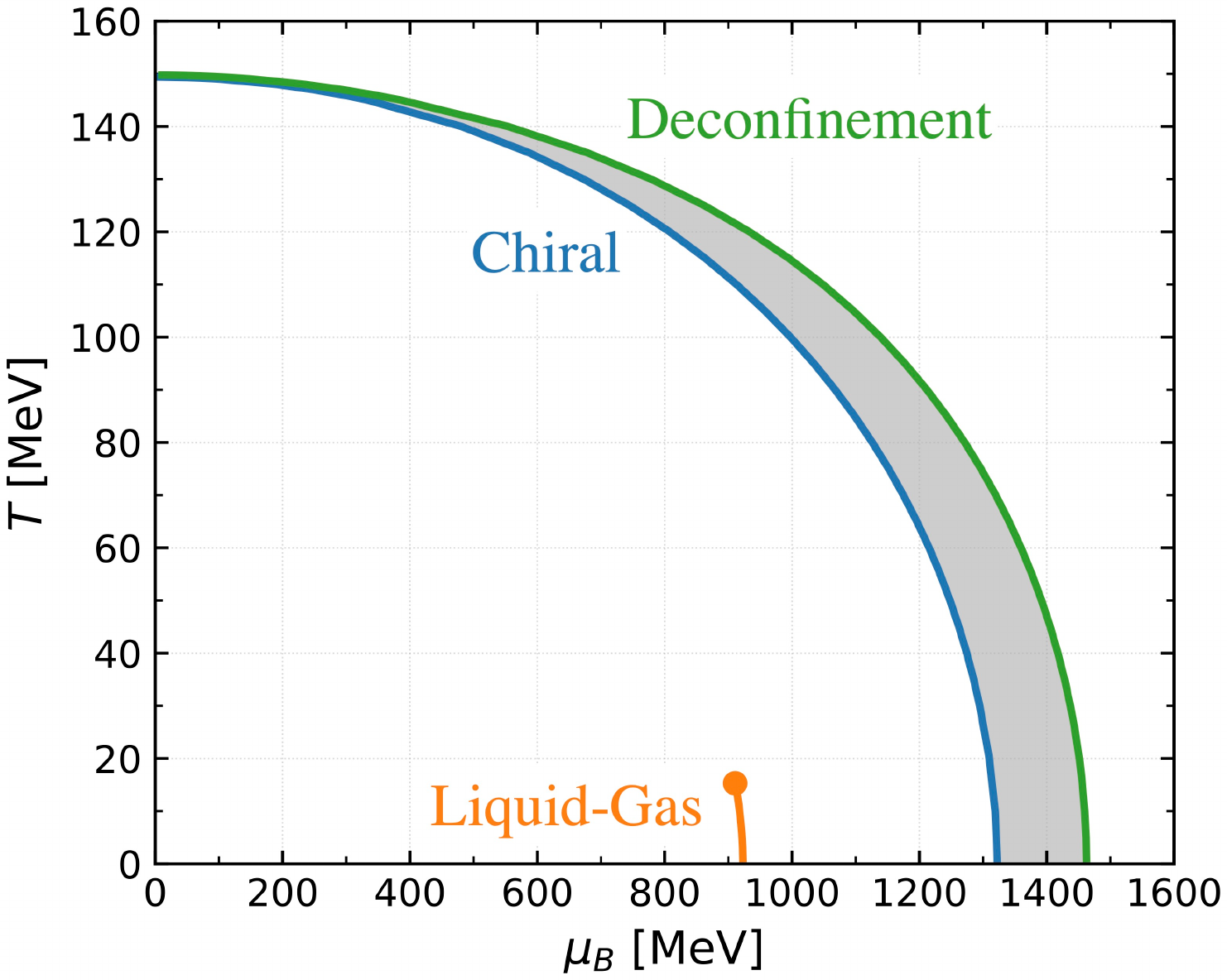}
    \caption{{Phase diagram predicted by our model. The green curve indicates the deconfinement phase transition, given by a first order transition between hadron and quark matter. The blue curve corresponds to the chiral phase transition within the quark sector, as detailed in Fig.~\ref{fig:condensados}. The light gray region between both curves is discussed in the main text. The small orange curve indicates the nuclear liquid–gas coexistence line in the hadron phase, which terminates at a critical point (orange dot).}}
    \label{fig:phase_diag}
\end{figure}

\subsection{Phase diagram}
\label{subsec:diagphase}

Analyzing the QCD phase diagram within the EVA–01 model with the chosen parametrization, we identify three distinct first-order phase transitions: the nuclear liquid-gas transition at low baryon densities, the deconfinement of hadronic matter into quarks, and a chiral-symmetry–restoring transition (see Fig. \ref{fig:phase_diag}). Their first-order character is evidenced in the $P–\mu_B$ plane by the emergence of multiple branches—corresponding to different solutions of the equations of motion—that cross at the transition points, producing discontinuities in the first derivatives of the Gibbs free energy. To select the thermodynamically favored phase, we adopt in all three cases the standard Maxwell construction, i.e., we choose the branch with the highest pressure at fixed baryonic chemical potential $\mu_B$. Our results show that these transitions remain first order over the entire temperature range investigated. We now turn to a more detailed discussion of these three transitions, outlining their key features and the methodology used to describe them.

At intermediate baryon chemical potentials and relatively low temperatures, we find a first-order phase transition within the hadronic sector—the analogue of the nuclear liquid–gas transition. Within the RMF formalism, which does not explicitly include nuclei as degrees of freedom, this transition manifests as a passage from vacuum to bulk nuclear matter at $\mu_B \approx 915~\mathrm{MeV}$. The resulting coexistence line terminates at a critical end point at $T_c^{\mathrm{LG}} = 15.3~\mathrm{MeV}$. This value is in excellent agreement with phenomenological constraints derived from multifragmentation analyses in low-energy heavy-ion collisions, which indicate a critical temperature of around $15~\mathrm{MeV}$~\cite{Fukushima:2011tpd, Fukushima:2013tpd, Buballa2005:nma}.

At higher chemical potentials and temperatures, the model describes the hadron-quark deconfinement transition. This transition is governed by the dynamics of the order parameter $\Phi$. Solving the equations of motion yields two distinct solution branches in the $P$-$\mu_B$ plane: a hadron-dominated phase, characterized by $\Phi \approx 0$, and a deconfined quark-matter phase, where $\Phi \approx 1$. A key feature of our parametrization is that these two phases are effectively pure, with no coexistence of hadron and quark degrees of freedom at a given chemical potential. The intersection of these two branches defines, via the Maxwell construction, a sharp first-order phase transition from the hadronic to the quark phase. This deconfinement boundary is shown as the green line in Fig.~\ref{fig:phase_diag} and remains first order over the entire range investigated, exhibiting no critical end point (CEP).

The third and final transition identified in our model is a first-order, chiral-symmetry–restoring transition that occurs entirely within the quark matter domain ($\Phi \approx 1$). It is identified by the intersection of two distinct branches of the quark phase: one characterized by a large chiral condensate and another where the condensate is significantly suppressed, signaling a partial restoration of chiral symmetry. However, it is crucial to note that this intra-quark transition appears in a chemical potential regime where the hadronic phase ($\Phi \approx 0$) remains the thermodynamically favored one, i.e., it possesses a higher pressure. Consequently, the chiral transition is thermodynamically subleading and does not manifest in the final, composite EoS. This is visually represented in Fig.~\ref{fig:phase_diag}, where the blue line corresponding to this transition lies entirely within the region governed by the hadronic phase.

To quantify the chiral transition, we calculate the light quark chiral condensate, $\langle \bar{q}q \rangle$, which is defined as \cite{Schwarz:2003tot, Jankowski:2013cci}
\begin{equation}
    \langle \bar{q}q \rangle = \langle \bar{q}q \rangle_0 + \frac{\partial \Omega}{\partial m_q }  \, ,
\end{equation}
where $\langle \bar{q}q \rangle_0$ is the vacuum condensate value, $\Omega = -P$ is the grand canonical potential, and $m_q$ is the current quark mass. We present results only for the light quark condensate $\langle \bar{u}u + \bar{d}d \rangle$; the strange condensate varies only smoothly at higher chemical potentials, as is generally observed in effective theories (see, for example, Fig.~$3.4$ and related discussion in Ref.~\cite{Buballa2005:nma}).

For the light $u$ and $d$ quarks, the vacuum condensate is fixed by the Gell-Mann-Oakes-Renner relation \cite{Chanfray:1994pdc, Schwarz:2003tot}:
\begin{equation}
    2 m_\pi^2 f_\pi^2 = - (m_u+m_d) \langle \bar{u}u + \bar{d}d \rangle_0 \,,
\end{equation}
where $m_\pi=138$~MeV is the pion mass and \mbox{$f_\pi=92.4$~MeV} is its decay constant. At finite temperature, we include the leading-order correction \cite{Gasser:1987lga, Agasian:2001gmo}
\begin{equation}
    \frac{\langle \bar{u}u + \bar{d}d \rangle_{0,T}}{\langle \bar{u}u + \bar{d}d \rangle_0} = 1 - \frac{T^2}{24 f_\pi^2} \, .
\end{equation}
Given that the scalar number density is defined as $\partial\Omega/\partial m^* = n_q^s$ (and $\partial m^*/\partial m = 1$), the condensate finally becomes
\begin{equation}
     \frac{\langle \bar{u}u + \bar{d}d \rangle}{\langle \bar{u}u + \bar{d}d \rangle_{0,T}} = 1 - \frac{(m_u+m_d)}{2 m_\pi^2 f_\pi^2 (1- \frac{T^2}{24 f_\pi^2})} (n_u^s + n_d^s)  \, .
\end{equation}

\begin{figure*}[t!]
    \centering
    \includegraphics[width=0.45\linewidth]{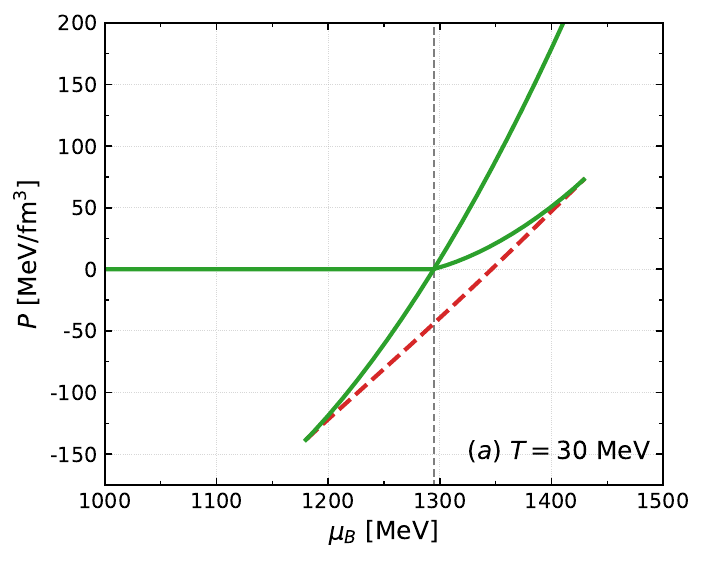}
    \includegraphics[width=0.45\linewidth]{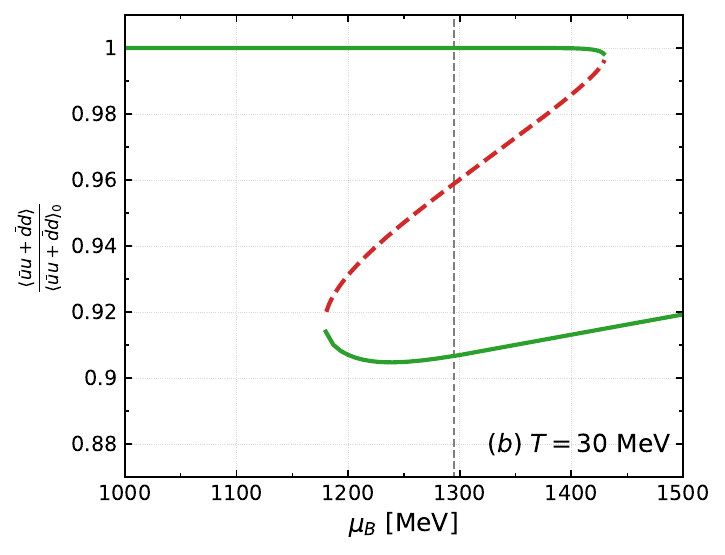}
    \includegraphics[width=0.45\linewidth]{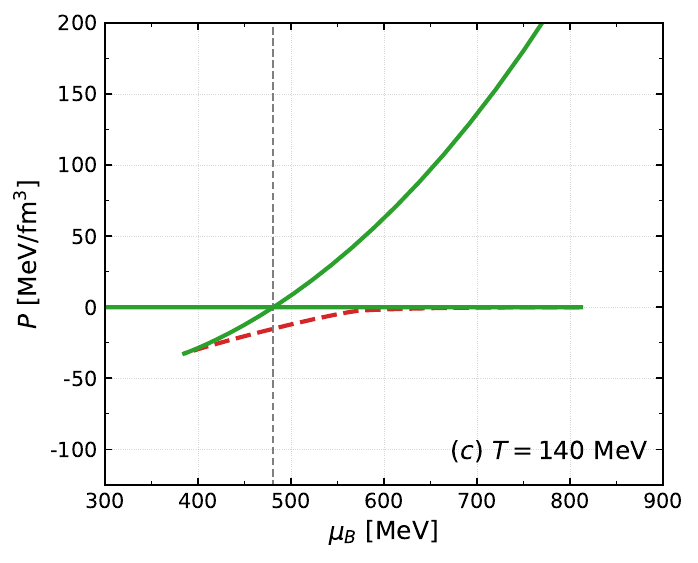}
    \includegraphics[width=0.45\linewidth]{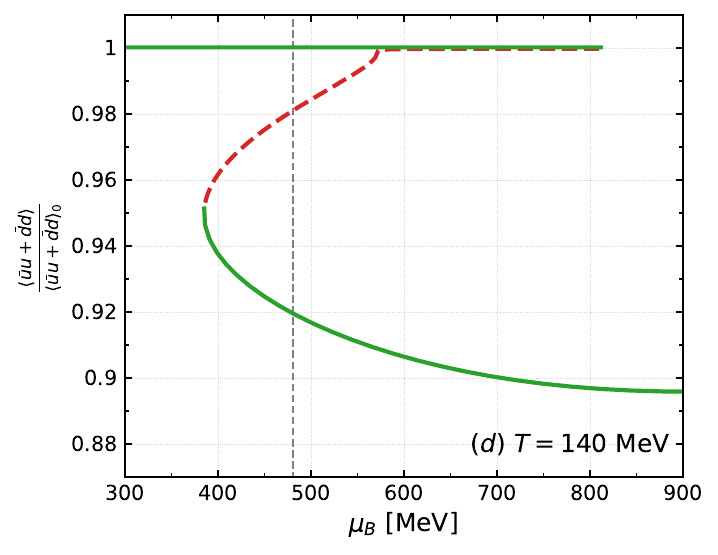}
    \caption{
     $P$–$\mu_B$ diagram, panels~$(a)$ and $(c)$, and light chiral condensate $\langle \bar{u}u + \bar{d}d \rangle$, panels~$(b)$ and $(d)$, for the quark phase at two representative temperatures, $T = 30$~MeV and $T = 140$~MeV. In panels~$(a)$ and $(c)$, the crossing of two stable branches (green curves) indicates a first-order phase transition. In panels~$(b)$ and $(d)$, the sharp drop in the chiral condensate at the same $\mu_B$ value confirms the partial restoration of chiral symmetry. The vertical dashed line marks the transition point in each case.}
    \label{fig:condensados}
\end{figure*}

This chiral phase transition characterization is illustrated in Fig.~\ref{fig:condensados} for two representative temperatures, $T = 30~\mathrm{MeV}$ and \mbox{$T = 140~\mathrm{MeV}$}. Panels~$(a)$ and $(c)$ display the $P$–$\mu_B$ curves, where the crossing of the two stable branches determines the location of the first-order transition. Panels $(b)$ and $(d)$ confirm the nature of this transition by showing a sharp drop in the chiral condensate at the exact same $\mu_B$ value. In all panels, the transition point is indicated by a vertical dashed line.

A key feature of our phase diagram, evident in Fig.~\ref{fig:phase_diag}, is that the chiral and deconfinement transitions do not occur simultaneously. This separation arises naturally from the presence of two distinct (approximate) order parameters —the chiral condensate for chiral symmetry and the Polyakov-like field $\Phi$ for deconfinement— that are only moderately coupled through the quark contribution to the thermodynamic potential. This structure is closely analogous to PNJL frameworks, where the effective potential comprises a chiral sector, a Polyakov sector, and a quark determinant that links them, yet does not fully lock their responses. Because the stationarity conditions for the condensate and for $\Phi$ react differently to changes in temperature and baryon chemical potential, the associated pseudocritical lines need not coincide. This mismatch can open an intermediate domain in which chiral symmetry is (partially) restored while matter remains confined—often discussed as a possible quarkyonic regime~\cite{McLerran:2007pod, Bluhm:2025qsi, Fukushima:2016tqs}. The degree of separation is, however, tunable: in entangled PNJL (ePNJL) models, introducing an explicit $\Phi$-dependence in the effective couplings tends to bring the two transitions closer~\cite{Sakai:2010ebd}; an analogous strategy in EVA–01 would be to strengthen the coupling between $\Phi$ and the effective masses. 

Furthermore, the specific functional form and parametrization of the Polyakov-like potential itself can alter the alignment of the two transitions. Conversely, adding a repulsive vector interaction in the quark sector is known to push chiral restoration to higher $\mu_B$ -and to move the CEP-, which can increase the relative misalignment with deconfinement depending on parameters. In the present parametrization of EVA–01, however, the hadronic branch remains thermodynamically favored up to the deconfinement boundary, effectively postponing the onset of chiral restoration in the equilibrium EoS. These considerations underscore the flexibility of the unified framework to probe the interplay between chiral symmetry restoration and deconfinement.

We now compare our results with established theoretical expectations. At $\mu_B = 0$, Lattice QCD simulations predict a smooth crossover, and many QCD-inspired models suggest the existence of a CEP at $(\mu_\mathrm{B, CEP}, T_\mathrm{CEP})$ where a first-order line begins~\cite{Stephanov:1998sot,Fischer:2019qaf}. Our model, predicting a first-order transition across the entire range, does not feature a crossover or a CEP. However, our results are compatible with those obtained using the concept of the \textit{Widom line}, which has been argued to be the remnant of the phase transition in the crossover regime (see, for example, Refs.~\cite{xu:2005rbt, simeoni:2010twl, ruppeiner:2012tgp, sordi:2024itc} and references therein). This compatibility is further strengthened by the fact that the predicted pseudo-critical temperature at zero chemical potential, $T_\mathrm{pc} = 149.8$~MeV, is in excellent agreement with lattice QCD estimates~\cite{Karsch:2001qma, Fukushima:2011tpd, Fischer:2021qpt}. Functional QCD methods also indicate that, if a CEP exists, it would likely occur at temperatures above \mbox{$T_\mathrm{CEP} \gtrsim 110$~MeV}~\cite{Gunkel:2021ltc, Gao:2021cps}.
Thus, despite this discrepancy, the model remains relevant for astrophysical applications, since temperatures \mbox{$T \sim 110 \, \mathrm{MeV}$} are not reached in those environments. In particular, because this work focuses on PNS applications where core temperatures typically remain below $50-60 \,\mathrm{MeV}$, the model's high-temperature limitations do not affect the validity of our astrophysical results.

It is worth noting that the features of the resulting phase diagram are achieved by a careful tuning of the model parameters. The value of $a_0$ was adjusted to obtain a pseudo-critical temperature $T_\mathrm{pc}$ closer to lattice QCD estimates, while the coupling $g_{\Phi b,q}$ was set to ensure a sharp separation between the hadron and quark phases, with no \textit{cross-contamination} across the entire range of temperatures and densities covered by the phase diagram. The modifications of parameters $a_1$ and $g_{\omega q}$ primarily affect the transition density, a crucial aspect for the astrophysical applications discussed in the following subsection.

%--------------------------------------------------------
\subsection{Astrophysics \label{subsec:astro}}
%--------------------------------------------------------

PNSs are formed in the aftermath of core-collapse supernova events as hot, dense compact objects gravitationally decoupled from the ejecta. Their subsequent evolution is complex. Initially, up to about $200$~ms after the core bounce, the PNS undergoes a rapid contraction, shrinking from a radius larger than $150$~km to less than $20$~km \cite{pons:1999eop,radice:2019ctg,burrows:2021ccs}. Following this, the star enters the quasi-stationary Kelvin-Helmholtz phase, which lasts for tens of seconds. This phase is dominated by two main thermal processes: first, a deleptonization stage, where trapped neutrinos diffuse outwards, heating the PNS core; and second, a global cooling stage, which concludes when the PNS becomes transparent to neutrinos and settles into a cold, catalyzed neutron star \cite{Camelio:2018eeo,pons:1999eop}.

To model this intricate evolution, we adopt three static snapshots that capture the thermodynamic conditions at key stages of PNS evolution. Based on previous studies that identify them as representative of the transient PNS state \cite{Burrows:1986tbo, Prakash:2001eoa, Mariani:2017ceh, Zheng:2025fmo}, these snapshots are:
\begin{itemize}
    \item \textit{Stage~$1$}: Represents the early PNS ($t \sim 1-2$~s). It is modeled as a core isentropic state with an entropy per baryon $s = \tilde{s}/n_B \approx 1$ (in units of $k_B$) and a high trapped lepton fraction ($Y_{l_e} = Y_e + Y_{\nu_e} = 0.4$), without muons present.
    
    \item \textit{Stage~$2$}: Corresponds to the subsequent, hotter phase ($t \sim 10-15$~s) attained after neutrino diffusion has significantly heated the core. This state is modeled as a core isentropic configuration with $s \approx 2$, where neutrinos have escaped ($Y_{\nu_e} = 0$) and muons are present.
    
    \item \textit{Stage~$3$}: Represents the final state after some minutes, a cold ($T=0$), catalyzed neutron star in full thermodynamic equilibrium.
\end{itemize}
The defining thermodynamic conditions for each stage are summarized in Table~\ref{table:thermo_snapshots}.

%%%%%%%%%   TABLE
\begin{table}[tb!]
\centering
\setlength{\tabcolsep}{5pt}
\renewcommand{\arraystretch}{1.2}
\begin{tabular}{cccccc}
\toprule 
Stage  & $s/k_B$ & $Y_{l_e}$ & $Y_{\nu_e}$ & $Y_\mu$ & Crust \\
\midrule
 1 & $\sim 1$ & $0.4$  & $\beta$ eq. & $0$ & $T=5$~MeV \\
 2 & $\sim 2$ & $\beta$ eq. & $0$ & $\beta$ eq. & $T=2$~MeV \\
 3 & $0$      & $\beta$ eq. & $0$ & $\beta$ eq. & $T=0$~MeV \\
\bottomrule
\end{tabular}    
\caption{Thermodynamic conditions defining the three evolutionary snapshots of a PNS. For stage~$1$, the electronic lepton fraction is fixed ($Y_{l_e} = Y_e+Y_{\nu_e} = 0.4$) to account for trapped neutrinos. In all other instances labeled `$\beta$ eq.', the corresponding value is determined by the condition of beta-equilibrium.} 
\label{table:thermo_snapshots}
\end{table}
 
For each snapshot, we construct the corresponding hybrid EoS. The core is described by our unified EVA-01 model, which yields a sharp, first-order hadron-quark phase transition. Contrary to the preceding analysis of the phase diagram, for these astrophysical applications we include the contribution from all relevant leptons ($e^{-}$, $\mu^{-}$, and $\nu_{e^-}$). The state of matter is determined by simultaneously solving the model's mean-field (gap) equations~\eqref{eq:motion} along with the conditions for $\beta$-equilibrium~\eqref{eq:chemeqb} and electric charge neutrality~\eqref{eq:charge_0}.

An important clarification regarding the inclusion of the crust for the outer layers must be made. EVA-01 is not capable of describing the behavior of matter in the low-density region of the crust. For this reason, independent models are used, and the interface between the crust and hadronic is set to occur at $n = 0.5\,n_0$. Furthermore, we adopt different crust EoSs to reflect the physical expectation that the crust cools monotonically, even as the core temporarily heats up \cite{Roberts:2012anc, Camelio:2018eeo, Pascal:2021mpn}. Specifically, we use the hot crust EoS from \citet{Dehman:2024iot} with $T=5$~MeV for stage~$1$ and $T=2$~MeV for stage~$2$. For the final cold configuration (stage~$3$), we employ the traditional BPS-BBP crust \cite{Baym:1971tgs, Baym:1971nsm}; although more modern models exist, the traditional BPS–BBP EoS is still widely used, as it remains consistent with current knowledge in the low-density regime (for more details regarding the use of the BPS-BBP crust and a comparison with other crust models see, for example, Refs.~\cite{fortin:2016nsr,canullan:2025ccc} and references therein).
It is important to clarify that the crust EoSs employed for the three stages are almost identical in the outer crust region, as \citet{Dehman:2024iot} used a $T=0$ approximation in this sector; in the inner crust, where finite temperature effects become relevant, the EoS crust used for the hot stage~$1$ is the stiffest, that for the cold stage~$3$ is the softest, and the hot stage~$2$ EoS crust lies in between for most of the inner crust density range.

%alternative theoretical frameworks to model the crust-hadron interface, see, for example, Refs.~\cite{fortin:2016nsr,canullan:2025ccc} and references therein)

Finally, with the complete hybrid EoS for each snapshot, we integrate the Tolman-Oppenheimer-Volkoff (TOV) equations to obtain the stellar structure, including the gravitational mass $M$ and the baryonic mass $M_B$ \cite{mariani:2019mhs}. Following the approach of \citet{Bombaci:1996tmm}, the analysis of the evolutionary path and stability of the PNS is then performed in the baryonic mass-gravitational mass plane, assuming the baryonic mass remains constant during the isolated evolution of the star \cite{Mariani:2017ceh}.

%%%%%%%%%%%    FIGURE 3
\begin{figure}[t!]
\centering
\includegraphics[width=0.99\linewidth]{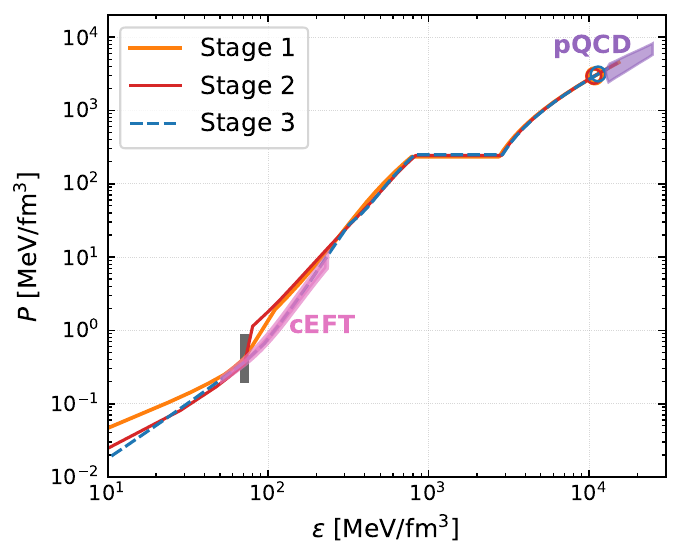}
\caption{Pressure-energy density relationship for the three PNS snapshots: stage~$1$ (orange), stage~$2$ (red), and stage~$3$ (blue, dashed). The shaded regions indicate the theoretical constraints from chiral Effective Field Theory (cEFT) at low densities \cite{Drischler:2021lma} and perturbative QCD (pQCD) at high densities \cite{Kurkela:2009cqm, Gorda:2018nop, Annala:2020efq}. A vertical gray line marks the approximate crust-core boundary at $0.5\, n_0$, and small circles denote the maximum central density reached in stable stellar configurations. Note that the logarithmic scale, necessary to display the wide range of values, obscures many of the differences between the three EoS, particularly in the transition plateaus and throughout the intermediate and high-density regions.}
\label{fig:p-eps}
\end{figure}

%%%%%%%%%%   FIGURE 4
\begin{figure*}[t!]
\centering
\includegraphics[width=0.99\linewidth]{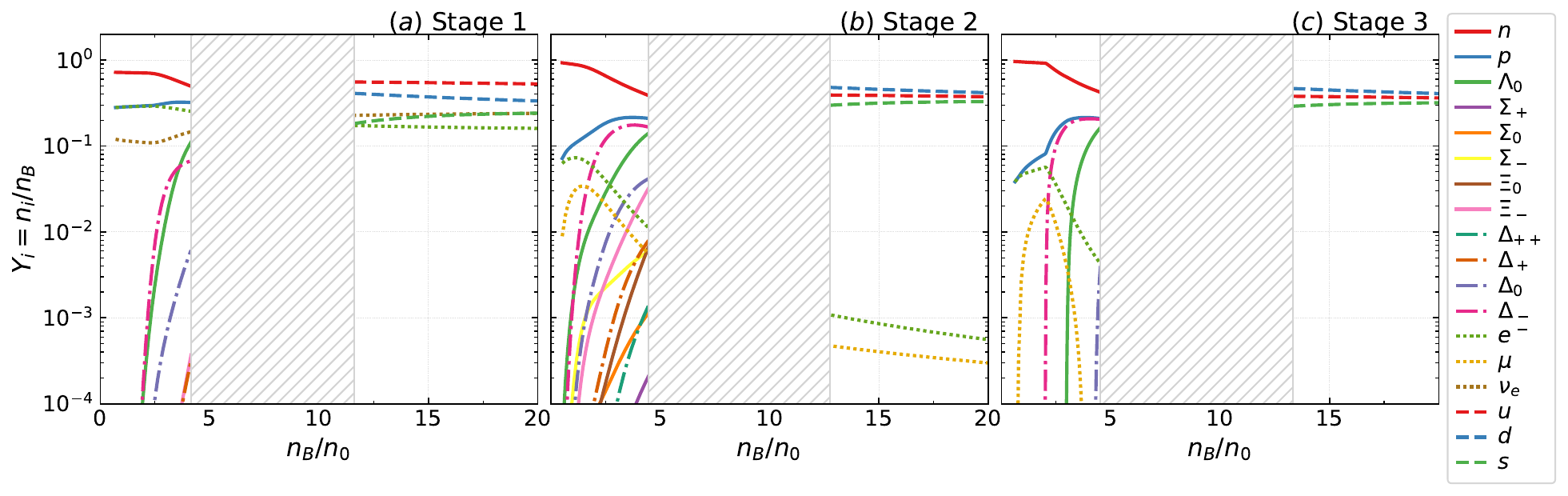}
\caption{Particle abundances ($Y_i$) as a function of normalized baryon number density ($n_B/n_0$) for three PNS snapshots: $(a)$ stage~$1$, $(b)$ stage~$2$, and $(c)$ stage~$3$. In each case, the hadron phase is separated from the quark phase by a first-order phase transition, indicated by the hatched region representing the density gap.}
\label{fig:abundances}
\end{figure*}

The EoS for the three evolutionary snapshots are shown in Fig.~\ref{fig:p-eps}.  It is important to note that only the cold EoS (stage~$3$) is expected to satisfy both cEFT and pQCD constraints. Indeed, the stage~$3$ model, which combines the BPS-BBP crust with the EVA-01 parametrization, is fully consistent with the low-density limit imposed by cEFT and with the high-density regime predicted by pQCD. In contrast, the hot EoSs for stage~$1$ and $2$ are stiffer, a direct consequence of their thermal contributions. All three curves exhibit a first-order hadron-quark phase transition, with pressures around $P_t \sim 220-240$~MeV/fm$^3$ and a large energy density gap of $\Delta \epsilon \sim 2000$~MeV/fm$^3$. As we will discuss later, the cores of the most extreme dynamically stable objects in our model reach densities of up to $n_B / n_0 \lesssim 40$. While extremely high, these densities do not reach the asymptotic regime where pQCD applies, as indicated by the small circles in Fig.~\ref{fig:p-eps}.

Figure~\ref{fig:abundances} shows the particle composition for the three evolutionary snapshots. A key feature across all stages is the sharp, first-order phase transition from hadronic to quark matter, which occurs at densities below $n_B/n_0 \lesssim 5$ and results in a large density gap of $\Delta n_B/n_0 > 5$. Our unified model self-consistently determines the presence or absence of particle species in each phase, and the proposed parametrization ensures that cross-contamination is negligible ($Y_i \lesssim 10^{-20}$).

The specific thermodynamic conditions of each snapshot lead to distinct compositions:
\begin{itemize}
    \item \textit{Stage~$1$:} The high trapped lepton fraction ($Y_{l_e} = Y_e + Y_{\nu_e} = 0.4$) results in significant electron and neutrino populations in both the hadron and quark sectors, which also enhances the proton abundance at lower densities and the light quark 
    $u$ abundance at high densities.
    \item \textit{Stage~$2$:} This hotter, neutrino-free state is characterized by high entropy, which allows for the thermal population of the full baryon octet and $\Delta$-resonances within the hadronic phase. The lepton fraction in both phases is also considerably larger compared to the cold scenario.
    \item \textit{Stage~$3$:} The final cold, catalyzed state shows a standard $npe$ composition at low densities, with muons and heavier baryons appearing as density increases. Notably, the $\Delta^-$ resonance becomes prominent, influencing the charge neutrality condition. The quark sector settles into a simple $uds$ composition with a negligible lepton presence.
\end{itemize}

\begin{figure}[t!]
\centering
\includegraphics[width=0.99\linewidth]{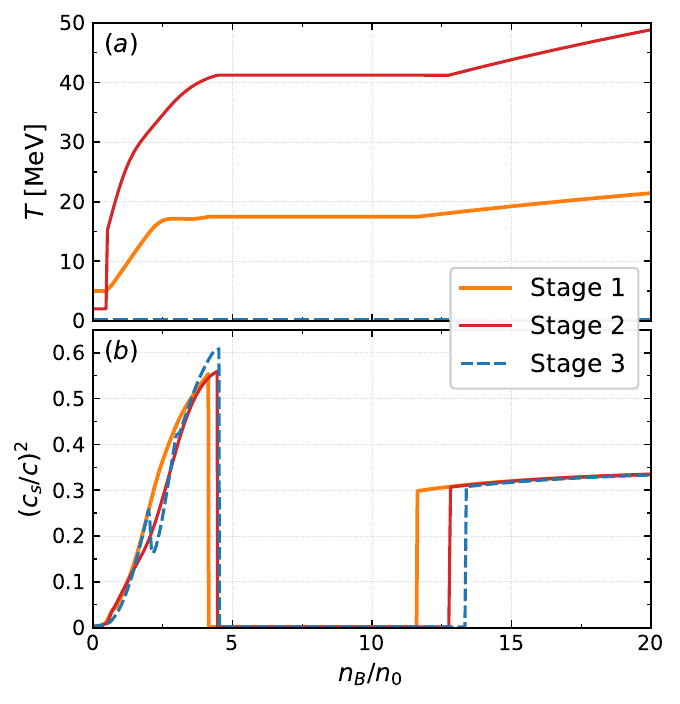}
\caption{Temperature $(a)$ and squared speed of sound, $(c_s/c)^2$ $(b)$, as a function of normalized baryon number density for the three PNS snapshots: stage~$1$ (orange line), stage~$2$ (red line), and stage~$3$ (blue, dashed line). Panel $(a)$ shows the isothermal crusts at low density and the thermal continuity at the phase transition for the hot stages. Panel $(b)$ illustrates the behavior of the speed of sound, including the convergence to the conformal limit ($(c_s/c)^2 = 1/3$) at high densities.}
\label{fig:temperature_cs2}
\end{figure}

Figure~\ref{fig:temperature_cs2}$(a)$ shows the temperature profiles as a function of normalized baryon number density for the three evolutionary snapshots. The structure of the profiles for the hot stages ($1$ and $2$) can be understood by analyzing three distinct regions. At low densities (approximately $n_B/n_0 \lesssim 0.5$), the temperature is constant. This isothermal region corresponds to the stellar crust, for which we assume a monotonic cooling from $T=5$~MeV in stage~$1$, through $T=2$~MeV in  stage~$2$, down to $T=0$~MeV in stage~$3$. For densities above the crust and up to the onset of the phase transition, the matter enters the isentropic hadron phase. In this region, the temperature generally increases with density as the matter is compressed. The transition to the quark phase occurs across a density gap, as illustrated in Fig.~\ref{fig:abundances}. The hadron-quark interface is constructed to maintain thermal equilibrium, which requires the temperature to be continuous across the phase boundary; otherwise, a net heat flow would occur. This condition results in transition temperatures of $T=17.6$~MeV for stage~$1$ and $T=41.5$~MeV for stage~$2$. A direct consequence of enforcing thermal continuity in this first-order phase transition is a discontinuity in the entropy density.  For instance, for stage~$1$ the entropy per baryon jumps from $s_H=0.94$ (hadron phase) to $s_Q=1.07$ (quark phase), while for stage~$2$ the jump is from $s_H=1.87$ to $s_Q=2.13$. This jump corresponds to the latent heat of the hadron-to-quark matter conversion. Finally, in the high-density quark phase, the temperature continues to rise with density, following a new isentropic path defined by a slightly higher entropy value than that of the hadronic phase.

In Fig.~\ref{fig:temperature_cs2}$(b)$, the squared speed of sound, $(c_s/c)^2$, shows distinct behaviors in the hadron and quark sectors. In the hadron phase, the speed of sound varies significantly with the thermodynamic conditions of each stage. The $T=0$ curve (stage~$3$) exhibits sharp changes associated with the abrupt appearance of new particle species. In contrast, the finite-temperature curves for stages~$1$ and $2$ are generally smoother, with the stage~$1$ EoS reaching the highest peak value for $(c_s/c)^2$ in this region.  In the quark phase, the speed of sound for all three stages converges toward the conformal limit of $(c_s/c)^2 = 1/3$, a behavior consistent with the pQCD constraint.

\begin{figure*}[t!]
\centering
\includegraphics[width=1.4\columnwidth]{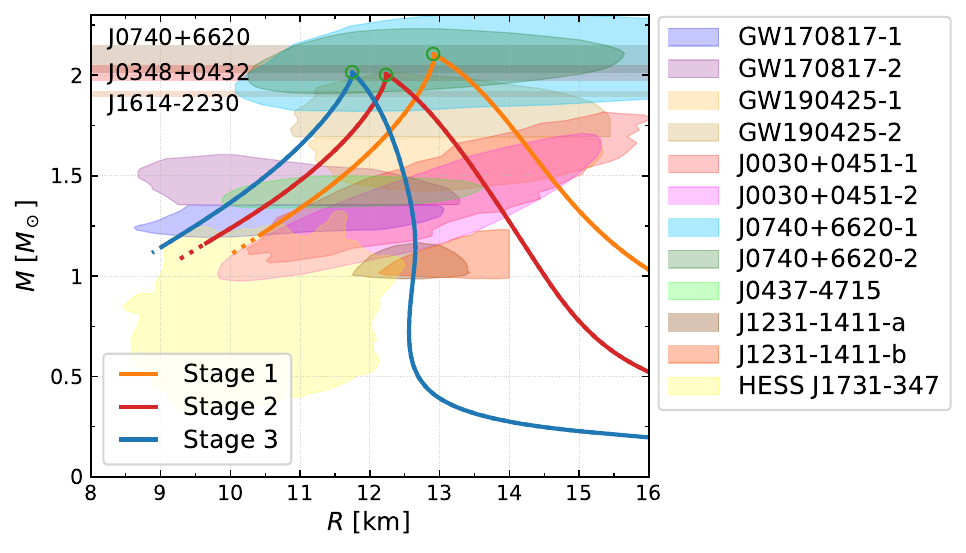}
\caption{Mass-radius ($M$-$R$) relationship for the three PNS snapshots: stage~$1$ (orange), stage~$2$ (red), and the final cold NS (stage~$3$, blue). Continuous lines represent dynamically stable stars, including the SSHS branch that emerges after the maximum mass peak. The green circle on each curve marks the appearance of the first configuration with a quark core. The plot is overlaid with modern astrophysical constraints from observations of high-mass pulsars (horizontal bands) \cite{Demorest:2010ats, Arzoumanian:2018tny,Cromartie:2020rsd, Fonseca:2021rma, Antoniadis:2013amp}, gravitational-wave events GW170817 and GW190425 \cite{Abbott:2017goo, Abbott:2017gwa, Abbott:2017moo, Abbott:2020goo}, X-ray timing of pulsars by NICER \cite{Riley:2019anv, Raaijmakers:2019anv, Miller:2019pjm, Riley:2021anv, Miller:2021tro, Choudhury:2024anv, Reardon:2024tns, salmi:2024anv}, and the analysis of the central compact object HESS~J1731$-$347 \cite{Doroshenko:2022asl}. The model for the final cold NS (stage~$3$) is compatible with all presented constraints.}
\label{fig:mraio}
\end{figure*}

The mass-radius ($M$-$R$) relations for the three evolutionary snapshots are presented in Fig.~\ref{fig:mraio}. The final cold NS sequence (stage~$3$) is consistent with the full set of modern astrophysical constraints shown. This agreement is achieved because the hadronic sector of \mbox{EVA-01} was adjusted to be compatible with low-mass constraints (except HESS J1731-347), while the maximum mass was specifically adjusted to reach $2.01 M_\odot$. This tuning was accomplished by modifying the model's $a_1$ parameter and activating the $\omega$-vector-meson interaction within the quark phase, which sets the hadron-quark phase transition at a density of $n_B/n_0 \sim 5$. Furthermore, our model is compatible with the challenging constraint of HESS J1731-347\footnote{We note that the inferred stellar parameters for this object are strongly sensitive to the assumed atmospheric composition.}. This is possible due to the slow hadron-quark conversion scenario \cite{Pereira:2018pte}, which produces a long branch of SSHSs after the maximum mass peak. The length of this branch is a direct consequence of the large energy density jump at the phase transition—a feature of our parametrization— which allows very compact and stable configurations \cite{lugones:2023ama, Mariani:2024cas, Laskos:2024hsi}.

The evolution from the hot stages of the PNS to the cold final NS shows a systematic contraction and a noticeable decrease in the maximum supported mass from stage~$1$ to stage~$2$, followed by an almost negligible increase from stage~$2$ to stage~$3$. The differences between the stages are driven by two main physical effects: thermal pressure and compositional changes due to neutrino trapping. In the core region, these effects can be described as follows:
\begin{itemize}
\item \textit{Thermal Effects:} The primary impact of finite temperature is to provide additional pressure support, but this effect is highly density-dependent. Although our calculations are performed using the full finite-temperature formalism, the physical effect can be understood through the Sommerfeld expansion. This approximation allows the total pressure to be separated into a cold component and a thermal correction, $P_{\text{total}} \approx P_{\text{cold}} + P_{\text{th}}$. While the cold pressure $P_{\text{cold}}$ grows very rapidly with density, the thermal pressure, $P_{\text{th}} \sim \mu_B^2 T^2$, increases much more moderately. Consequently, the fractional contribution of thermal pressure, $P_{\text{th}}/P_{\text{cold}}$, is significant at low to moderate densities but becomes negligible at the ultra-high densities found in the cores of maximum-mass stars. This density-dependent behavior directly explains the features seen in Fig.~\ref{fig:mraio}. A star's radius is highly sensitive to the pressure in its outer core. The substantial thermal pressure boost in this region effectively ``inflates'' the star, leading to the significantly larger radii observed for stages~$1$ and $2$ compared to stage~$3$. Conversely, a star's maximum mass is dictated by the EoS at the highest densities. In this regime, $P_{\text{th}}$ is only a minor perturbation on the dominant $P_{\text{cold}}$. 

However, finite temperature also introduces a crucial competing effect: the thermal excitation of heavy baryonic states, such as hyperons and $\Delta$ resonances, as seen in Fig.~\ref{fig:abundances}. At $T=0$ (stage~$3$), these heavy particles only appear at very high densities where the nucleon chemical potential exceeds their effective mass. In contrast, at the high temperatures of stage~$2$, the particle distributions are thermally smeared, allowing for a significant population of these states to be excited. The impact of this effect is highly density-dependent. At the lower densities that determine the star's radius ($n_B \sim 1-2\,n_0$), the nucleon chemical potential is still far below the effective mass of these heavy particles. Consequently, their thermal population is exponentially suppressed and the associated softening of the EoS is negligible. In this region, the stiffening from $P_{\text{th}}$ is the dominant thermal effect, which explains why the radii of the hot stars are significantly larger. Conversely, as density increases towards the values found in the cores of massive stars, the nucleon chemical potential approaches the heavy baryon mass thresholds. This makes the thermal excitation of hyperons and Deltas much more efficient, and their population grows rapidly. The appearance of these new degrees of freedom provides a significant softening of the EoS that becomes more pronounced with increasing density. At the extreme densities relevant for the maximum mass, this softening from the diverse baryonic population ultimately outweighs the stiffening from thermal pressure, explaining why $M_{\text{max}}$ for stage~$2$ is slightly smaller than for stage~$3$.
    
\item \textit{Neutrino Trapping (stage~$1$):} In stage~$1$, neutrino trapping fundamentally alters the matter's composition and pressure, making the EoS significantly stiffer. This compositional stiffening arises from two primary mechanisms. First, the condition of a high, fixed lepton fraction ($Y_{l_e}=0.4$) ensures a large population of both trapped neutrinos and electrons, as relativistic fermions exert a significant pressure component ($P_e + P_{\nu_e}$), that is absent in the later stages. 

Second, and critically, the high proton fraction forced by beta-equilibrium drastically alters the baryonic composition by strongly suppressing the appearance of heavy baryons, which, as previously discussed, soften the EoS. This suppression in stage~$1$ occurs for two complementary reasons. Primarily, the proton-rich environment keeps the neutron chemical potential, $\mu_n$, below the energy threshold required for hyperon creation up to much higher densities than in stage~$2$. Additionally, the temperature in stage~$1$ is considerably lower than in stage~$2$ (e.g., a transition temperature of $\approx 18$~MeV vs. $\approx 42$~MeV), which further reduces the thermal energy available to excite these massive states. As a result of both, a higher chemical threshold and a lower temperature, the thermal population of hyperons and Deltas is rendered negligible. The combination of a large direct lepton pressure and the suppression of hyperonic softening makes the stage~$1$ EoS significantly stiffer than that of stage~$2$. This has profound consequences for the entire sequence of stellar configurations, shifting the stable hybrid star branch for stage~$1$ to significantly higher masses and radii, as seen in Fig.~\ref{fig:mraio}.
\end{itemize}

%%%%%%%%%%   FIGURE 5

%%%%%%%%%%%%% FIGURE 6

Besides the core phenomena, the crust effects also contribute to the evolution of the PNS structure. Due to the low density of the crust, its impact on the total mass is not significant. However, the determination of radius, $R$, of the stellar configuration obtained by solving the TOV equations, relies on the boundary condition $P(R)=0$ (corresponding to the region where the crust dominates). For this reason, the specifics of the crust produce a significant impact on the total radius of the star (for more details on the impact of the crust layer on the mass and radius see Refs.~\citep{fortin:2016nsr, canullan:2025ccc}). 
The use of the hotter crust for stage~$1$ ($T=5$~MeV) produces a noticeable increase in the radius values compared to the stage~$2$ case ($T=2$~MeV), which in turn generates larger radii than stage~$3$ case (BPS-BBP cold crust). This result is in agreement with the original hot-crust work by \citet{Dehman:2024iot}. 

In summary, the combined effects of thermal pressure and composition dictate the evolutionary path of the PNS in the mass-radius diagram. This results in a continuous decrease of the stellar radius, and an initial strong drop (from stage~$1$ to $2$), followed by a quasi-stabilization (from stage~$2$ to $3$) of the maximum supported mass. The contraction of PNSs across all stages is in agreement with hydrodynamical simulations, which also predict a monotonic contraction throughout the entire PNS evolution~\cite{Camelio:2018eeo, Pascal:2021mpn, Heinlein:2023pns}.

\begin{figure}[t!]
\centering
\includegraphics[width=0.99\linewidth]{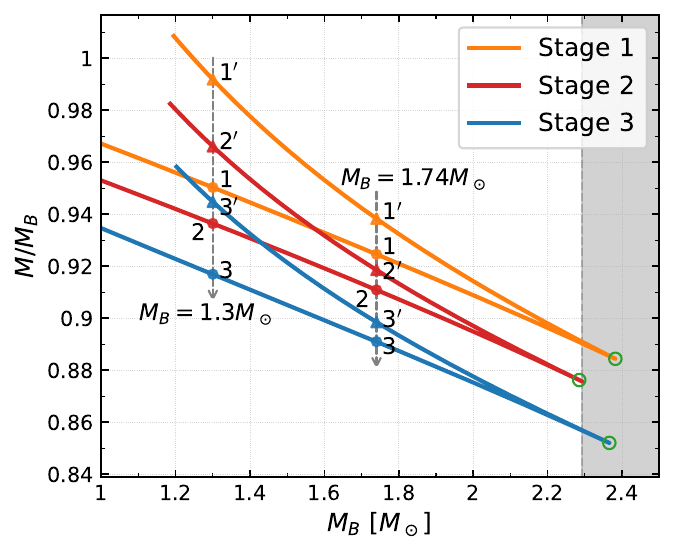}
\caption{Relation between the gravitational-to-baryonic mass ratio ($M/M_B$) and the baryonic mass ($M_B$) for the three PNS snapshots. All configurations shown on the curves are dynamically stable, including those on the backward-bending SSHS branches. The green dots on each curve denote the onset of a quark core, and the gray shaded region highlights the evolutionary instability window. The vertical arrows illustrate two evolutionary scenarios, where points on the hadronic branch are marked by numbers (not-prime) and their counterparts on the SSHS branch are marked by primed numbers. See the main text for a detailed analysis.}
\label{fig:mbmg}
\end{figure}

The relationship between the gravitational mass ($M$) and the baryonic mass ($M_B$) provides a powerful tool to analyze the evolutionary path of an isolated PNS. Fig.~\ref{fig:mbmg} shows this relationship for the stable configurations in our three evolutionary snapshots, with the $y$-axis normalized as $M/M_B$ for clarity. In the absence of significant fallback accretion, a PNS evolves at a constant $M_B$. Since the star radiates energy as it cools and deleptonizes, its gravitational mass must decrease over time, meaning that any physically allowed evolutionary pathway must proceed vertically downward in this diagram. For baryonic masses $M_B \gtrsim 1.2 M_\odot$, two families of stable solutions exist: a hadronic branch, which represents the global minimum of gravitational mass for a given $M_B$, and a hybrid SSHS branch, which is a local minimum. In the absence of large perturbations, a star is expected to remain in its initial potential well. This implies a smooth, quasi-static evolution: a star born on the hadronic branch would follow the $1 \rightarrow 2 \rightarrow 3$ track, while a star born as a SSHS would evolve along the $1' \rightarrow 2' \rightarrow 3'$ track.

However, the PNS environment is prone to significant instabilities. Large-scale convection, known to occur during the early stages of evolution, could act as a trigger for a jump between these two local minima.  A conversion  from the hybrid to the hadronic branch is always energetically favorable, as the hadronic state is the global energy minimum. A more speculative scenario is the conversion from a hadronic to a SSHS configuration. For a high-mass PNS (e.g., $M_B = 1.74 M_\odot$), a star in the hadronic state (point 1) may have energetically accessible hybrid states in its future evolution (e.g., point 2', since $M_{2'} < M_1$). If a sufficiently large perturbation triggers a complex hydrodynamic phase, the star could settle into a new quasi-static equilibrium once this dynamic phase subsides, possibly on the hybrid SSHS branch. This conversion path is less likely for lower-mass stars (e.g., $M_B = 1.3 M_\odot$) due to the interleaving of the branches. For a hadronic star at state 1, the first energetically accessible hybrid state might belong to a much later thermal epoch (e.g., state 3'). By the time the star cools to that stage, it will likely have already evolved to a lower-energy hadronic state (such as point $2$ or $3$), from which the transition to the hybrid branch is no longer possible. While highly speculative, this analysis of potential transitions illustrates a rich variety of possible evolutionary paths and underscores the need for more detailed numerical simulations.

An additional noteworthy feature in Fig.~\ref{fig:mbmg} is that the maximum supported baryonic mass for stage~$2$ is lower than for both stage~$1$ and stage~$3$. This implies the existence of an unstable evolutionary window for the most massive PNSs. A PNS born with a baryonic mass in the range $M_{B,\text{max}}(\text{stage~$2$}) < M_B < M_{B,\text{max}}(\text{stage~$1$})$ would be stable during its initial phase but would inevitably collapse into a black hole upon reaching the thermodynamic conditions of stage~$2$. Consequently, within an isolated evolutionary scenario, no cold neutron star can form with a baryonic mass greater than the stability limit set by stage~$2$ ($M_B \approx 2.3 M_\odot$ in our model), which corresponds to a gravitational maximum mass of $\sim 2.007 M_\odot$. Therefore, explaining the existence of observed massive NSs with gravitational masses~$\gtrsim 2 M_\odot$ would, in the context of the present EoS, require a post-PNS mass accretion phase, for instance through late-time fallback accretion or mass transfer in a binary system.

\section{Conclusions}\label{sec:conclus}

This work presents an effective model for a hybrid EoS that unifies hadron and quark degrees of freedom within a single Lagrangian. Our model is built upon the DDRMF framework (which includes density-dependent couplings for the $\rho$ meson) and, to achieve deconfinement, we adopt an approach similar to the CMF model, implementing a scalar field, $\Phi$, that drives the hadron-quark transition. The Lagrangian includes the full baryon octet, $\Delta$ resonances, $u,d,s$ quarks, and leptons, which interact via scalar ($\sigma, \sigma^*$), vector ($\omega, \phi$), and isovector ($\rho$) meson fields, along with the deconfinement field $\Phi$.

The hadron sector parametrization ensures that our model satisfies nuclear saturation properties, reproduces the liquid-gas phase transition, and meets constraints from both cEFT and observations of low-mass NSs. For the quark sector, we introduced a new parametrization for the quark-meson couplings and the $\Phi$ potential. While this reproduces reasonable deconfinement and chiral restoration curves in the phase diagram, a key limitation is the absence of a CEP and a crossover region at high temperatures. 

Applying our unified EoS, we modeled the evolution of PNSs through three representative thermodynamic snapshots, from their early hot and lepton-rich phases to their final states as cold, catalyzed NSs. This approach allowed us to study the impact of thermal effects and composition on the stellar structure. Our key findings are:

\begin{itemize}
    \item The final cold NS sequence (stage~$3$) is fully consistent with current astrophysical constraints, including observations of high-mass pulsars \mbox{($M_{\text{max}} \geq  2.01\,M_{\odot}$)}, GW events, and X-ray timing data. The model also accommodates the low-mass, high-compactness object HESS J1731-347 via the SSHS branch.

    \item The PNS evolution shows a systematic contraction as the star cools. The initial, lepton-rich stage~$1$ is the stiffest due to both direct lepton pressure and the suppression of heavy baryons, resulting in larger radii and a higher maximum mass.

    \item In the subsequent hot, neutrino-free phase (stage~$2$), the high temperature leads to a significant thermal population of hyperons and $\Delta$ resonances. The appearance of these new degrees of freedom softens the EoS at high densities, an effect that counteracts the thermal pressure, resulting in a maximum gravitational mass for this stage that is slightly lower than that of the final cold star.

    \item Following results from dynamical simulations of PNSs (see, for example, Refs.~\citep{Roberts:2012anc, Camelio:2018eeo, Pascal:2021mpn} and references therein), we adopt sequentially colder crusts (in contrast to the non-monotonic temperature evolution of the core) within our  snapshot-based construction. This treatment allows our model to reproduce the consistent contraction of the PNS throughout its evolution. It is worth noting that this crust modeling is neglected in most of the PNS snapshots studies, which typically show fluctuating radii across stages~\cite{Shao:2011eop,Mariani:2017ceh, Malfatti:2019hqm,Issifu:2024fuw, Ghosh:2024etm, Zheng:2025fmo, Kumar:2025com}.

    \item The analysis in the gravitational-baryonic mass plane reveals an instability window determined by the maximum baryonic mass of stage~$2$ (\mbox{$M_B \approx 2.3 M_\odot$}, corresponding to a maximum gravitational mass of $\sim 2.007 M_\odot$). A PNS born with a baryonic mass exceeding this limit would initially be stable but would inevitably collapse into a black hole during its thermal evolution. \citet{Kumar:2025com} speculates on the possible signatures of such delayed black hole formation triggered by early PNS destabilization.

    \item The coexistence of hadronic and SSHS branches for a given baryonic mass suggests complex evolutionary pathways. While quasi-static evolution would 
    preserve the star on its original branch,
    the turbulent PNS environment could trigger transitions between these states, highlighting the need for detailed dynamical simulations. In particular, unlike our results, Ref.~\cite{Issifu:2024fuw} suggests that during the early high-lepton stages, the EoS remains purely hadronic and the hadron-quark phase transition is only favored once  neutrinos escape. To establish potential smoking guns among models, it would be interesting to investigate  observational signatures of such delayed phase transition occurring seconds after PNS formation. 
\end{itemize}

Compared to the CMF model, which also uses a scalar field $\Phi$ to drive deconfinement \cite{Kumar:2024mna}, EVA-01 relies on similar principles, drawing inspiration from CMF's treatment of deconfinement and effective masses. While the CMF model successfully reproduces a CEP and a crossover region in the QCD phase diagram, EVA-01 does not yet exhibit such features. On the other hand, our model shows improved agreement with cEFT constraints and remains consistent with observations of massive pulsars and gravitational-wave events such as GW170817, even without invoking the SSHS hypothesis. Additionally, the two models differ in their high-temperature particle content: CMF allows for phase cross-contamination (i.e., quarks in the hadron phase and vice versa) \cite{Peterson:2022cse}, whereas the tuned couplings in EVA-01 are  adjusted to maintain phase purity across the full temperature range studied.

To sum up, at $T=0$, our model satisfies all current constraints from both nuclear physics and astrophysics. In the finite-temperature regime, while the resulting phase diagram is qualitatively reasonable, the absence of a CEP and crossover region remains an open issue for future development.

In closing,  the unified hadron–quark model EVA-01 establishes a coherent framework that incorporates essential microphysics while complying with the most recent nuclear and astrophysical constraints. Future theoretical work will focus on refining the model's core features. This includes a detailed investigation of the interplay between the deconfinement and chiral restoration transitions, as well as the inclusion of diquark pairing to describe color-superconducting phases. A major challenge will be to identify the conditions for incorporating a CEP and a crossover region while preserving essential features such as the absence of phase cross-contamination. Moreover, as mentioned in Section~\ref{sec:intro}, recent studies have suggested that the hadron-quark transition could undergo a crossover transition even in the low temperature, high density regime, challenging the traditional picture of a sharp first-order transition. On the other hand, we  plan to include the crust region within the unified scheme of EVA-01 in a future project; as the physics of the crust involves complexities beyond the mean-field description used here, it remained beyond the objectives of the present work. 
These future developments will, in turn, enable a broad range of astrophysical applications, including the study of HSs, the stability of SSHS, and their distinctive observational signatures.

\begin{acknowledgments}
We thank W.~Spinella and F.~Weber for developing the zero-temperature DDRMF-SW4L parametrization, which forms the basis of the hadron sector used in this work. We also thank R.~Kumar and V.~Dexheimer for the discussions during our CMF collaboration, through which we learned the implementation of the Polyakov‑like potential and its coupling to effective masses for the extension of our model to the quark sector. The authors thank N.~Scoccola for helpful discussions on the use of total and partial higher-order thermodynamic derivatives and their interpretation within effective models. M.~M. thanks C.~Dehman for kindly providing the finite-temperature crust EoSs used in this study, which helped shape important aspects of the approach adopted for PNS. M.~G.~O. thanks A.~Fantina, F.~Gulminelli, and N.~Chamel for the clarifications and comments on the crust EoS at finite temperature. M.~O.~C., M.~M., M.~G.~O. and I.~F.~R.-S. acknowledge UNLP and CONICET (Argentina) for financial support under grants 11/G187 and PIP-0169. G.L. acknowledges the financial support from the Brazilian agency CNPq (grant 316844/2021-7).
\end{acknowledgments}

% Produces the bibliography via BibTeX.
\bibliography{biblio}

\end{document}